\documentclass[conference]{IEEEtran}

\usepackage{xspace}
\usepackage{listings}
\usepackage{colortbl}
\usepackage{tabularx}
\usepackage{pgfplots}
\usepackage{pifont}
\usepackage[group-separator={,}]{siunitx}
\usepackage{diagbox}
\usepackage{multirow}
\usepackage{hyperref}
\usepackage{caption}
\usepackage{graphicx}
\usepackage[htt]{hyphenat}
\usepackage{cite}

\pgfplotsset{width=0.47\textwidth,compat=1.9}

\newcommand{\sysname}{Tinbergen\xspace}
\newcommand{\framework}{Software Ethology\xspace}
\newcommand{\gcc}{\texttt{gcc 7.5.0}\xspace}
\newcommand{\clang}{\texttt{clang 6.0.0}\xspace}

\newcommand{\accuracy}{$.779$\xspace}
\newcommand{\aarchaccuracy}{$.811$\xspace}
\newcommand{\improve}{$25\%$--$53\%$\xspace}
\newcommand{\diagaccuracy}{$.893$\xspace}
\newcommand{\offdiagaccuracy}{$.766$\xspace}
\newcommand{\labeltime}{$13.0$ CPU minutes\xspace}
\newcommand{\dtreetime}{$24.3$ CPU hours\xspace}

\newcommand{\treesize}{$855.9$ KB}
\newcommand{\obfusimprove}{$39.3$\%\xspace}

\newcommand{\valgrindloc}{$3,848$\xspace}
\newcommand{\pyloc}{$2,392$\xspace}

\newcommand{\dtreecount}{24\xspace}
\newcommand{\evalcount}{5\xspace}

\newcommand{\coreutils}{\texttt{coreutils-8.32}\xspace}

\lstdefinelanguage
[x64]{Assembler}     
[x86masm]{Assembler} 
{morekeywords={CDQE,CQO,CMPSQ,CMPXCHG16B,JRCXZ,LODSQ,MOVSXD,ANDQ,CMPQ, %
POPFQ,PUSHFQ,SCASQ,STOSQ,IRETQ,RDTSCP,SWAPGS,MOVQ,TESTB,ADDQ,PUSHQ,CMPB,INCQ, %
rax,rdx,rcx,rbx,rsi,rdi,rsp,rbp,dil, %
r8,r8d,r8w,r8b,r9,r9d,r9w,r9b, %
r10,r10d,r10w,r10b,r11,r11d,r11w,r11b, %
r12,r12d,r12w,r12b,r13,r13d,r13w,r13b, %
r14,r14d,r14w,r14b,r15,r15d,r15w,r15b}} 

\definecolor{mygreen}{rgb}{0,0.6,0}
\definecolor{mygray}{rgb}{0.5,0.5,0.5}
\definecolor{mymauve}{rgb}{0.58,0,0.82}
\author{\IEEEauthorblockN{Derrick McKee}
\IEEEauthorblockA{Purdue University\\mckee15@purdue.edu}
\and
\IEEEauthorblockN{Nathan Burow}
\IEEEauthorblockA{Purdue University\\nburow@purdue.edu}
\and
\IEEEauthorblockN{Mathias Payer}
\IEEEauthorblockA{EPFL\\mathias.payer@nebelwelt.net}
}

\begin{document}
    \title{\framework: An Accurate, Resilient, and Cross-Architecture
    Binary Analysis Framework}

    \maketitle

    \begin{abstract}
        Determining the semantic similarity of functions is a key task in
        reverse engineering that heretofore has been hindered by inaccuracy
        and/or high labor costs.
        \framework addresses inaccuracy and analysis cost by abstracting
        semantic behavior as classification vectors of program state changes.
        Our framework executes functions with a specified input state,
        leveraging these vectors as unique fingerprints for semantic
        identification.
        Existing binary analyses determine function similarity via code
        measurements, and suffer from high inaccuracy when classifying
        functions from compilation environments different from their training
        source.
        Since \framework does not rely on code measurements, it withstands broad
        changes in compiler, compiler version, optimization level, or even
        different implementations of equivalent functionality.
        As a key feature, classification vectors for one architecture are
        transferable to other architectures with minimal effort.

        \sysname, our prototype \framework implementation, deploys a fuzzer in
        a virtual execution environment to generate the classification vectors.
        Evaluating \sysname as a semantic function identifier for \coreutils
        functions, we achieve a high \accuracy average F-Score.
        Compared to the state-of-the-art \emph{BLEX} and \emph{IMF-SIM}
        frameworks, \sysname is \improve more accurate when identifying
        functions in binaries generated from differing compilation environments.
        We demonstrate that program state changes are versatile semantic
        identifiers, by achieving similarly high accuracy rates in
        purposefully obfuscated code, as well as when identifying functions
        in \texttt{AArch64} binaries using classification vectors generated
        from \texttt{x64} binaries.
        Finally, we show that \sysname scales to large binaries, by performing
        an evaluation on semantic identification accuracy for \texttt{libxml2},
        \texttt{libpng}, and \texttt{libz}, which are among the largest
        shared libraries distributed with Ubuntu.

    \end{abstract}


    \section{Introduction} \label{sec:intro}

    Semantic binary analysis --- the act of determining a function's
    ``purpose'' within a binary --- has applications in many research and
    engineering areas.
    In software forensics, identical code across different binaries highlights
    plagiarism or copyright violations~\cite{obfus_bin_sim}.
    Identical semantic code in a single binary is redundant, and can be removed
    by software engineers to limit binary bloat~\cite{detect_code_clones}.
    Determining how a new malware affects computer systems often starts with
    reverse engineering the semantic behavior of (new) functions present in the
    binary~\cite{rid2015attributing}.
    Malware authors often make slight changes from existing malware to create a
    new strain in an effort to evade detection, and analysts use
    semantic similarity (i.e., how close in functionality one function is to
    another) to deduce lineages in these newly discovered
    strains~\cite{auto_lineage,bayer2009scalable}.
    Attackers can find vulnerabilities in closed source software by performing
    differential semantic analysis on pre- and post-patched
    binaries~\cite{auto_patch_exploit}.

    As difficult as it is to determine a function's semantic behavior from
    source~\cite{cc_finder, cp_miner, deckard, auto_mining, semantic_clones}, it is
    even \emph{more} difficult from a compiled binary~\cite{meng2016binary}.
    Inferring semantics from code is challenging as the same source-level semantics
    can be implemented through vastly different code, and there is no direct
    relationship between the two.
    At the binary level, the analyst must first determine the set of changes
    to program state a function can conduct (e.g., what computations the
    function performs, or what memory addresses are accessed).
    Based on the program state changes the analyst can then infer semantic
meaning,
and eventually build a whole
    program understanding from how the semantic pieces fit together.
    Consequently, whole-program analysis at the binary level remains a challenge
    because manual analysis is time-consuming, and automated
    solutions~\cite{blanket, stat_sim, in_memory_fuzzer, bindiff, ida} are
    inaccurate or incomplete.

    Automated binary analysis solutions~\cite{shoshitaishvili2016sok} all
    measure properties of binary code (e.g., order and type of
    instructions~\cite{in_memory_fuzzer}, memory locations
    accessed~\cite{blanket, alphadiff}, or control flow~\cite{pewny2015cross}),
    comparing the behavior of functions based on the similarity of
    implementation artifacts.
    These solutions assume that code similarity approximates function semantic
    similarity, but machine code can vary while still preserving semantics.
    We demonstrate that program state modifications serve as a better, more
    stable semantic function identifier.
    Program state change as a function identifier relies on the fact that semantic
    behavior is stable across compilations, environments, or implementations.
    Thus, program state change provides an ideal fingerprint, as it is impervious to
    compilation environment diversity or information loss.
    Code measurement approaches, conversely, are susceptible to at least one of
    these complicating factors.

    In this paper, we present \framework, an approach to precise binary semantic
    analysis.
    Instead of relying on measuring code properties of a function, \framework
    abstracts functions into characteristic sets of inputs and corresponding
    program state changes.
    The core idea of \framework is to observe and identify the behavior or
    character of functions instead of the underlying code, and then use the
    observed behavior as a unique function identifier.
    We draw an analogy between the analysis of function semantics and the biological
    concept of ethology (from the Greek \emph{ethos} and \emph{logia}, meaning
    the \emph{study of character}).
    A major topic in ethology is the study of reliably predictable animal
    responses to the presence of known environmental stimuli.
    Likewise, semantically similar functions reliably make predictable changes to
    program state for given initial states.
    Those state changes can be measured and saved for later use in identifying
    unknown functions in a stripped binary, irrespective of the compilation
    environment.

    As a proof of concept for \framework, we implement \sysname\footnote{Named
    for Nikolaas Tinbergen, the founder of biological ethology}, a dynamic,
    fuzzer-inspired binary analysis tool.
    \sysname adapts mutational, coverage-guided fuzzers to discover a subset
    of a function's unique set of inputs and measurable program state changes
    (referred to as Input/Output Vectors, or \emph{IOVecs}).
    \sysname organizes the discovered IOVecs into a searchable tree, from which
    unknown functions can be classified as a previously analyzed function, or as
    genuinely unique functionality, which should be the focus of manual analysis.
    Additionally, \sysname identifies the course-grained layout of input
    and global structures used by IOVec exercised code, including
    a minimal size and which bytes are pointers.

    We evaluate \sysname on accuracy amid varying compilation environments, a
    task existing works find difficult yet is crucial for automated reverse
    engineering.
    We measure accuracy by identifying functions in the \coreutils application
    suite, and find that \sysname achieves a high \accuracy average
    accuracy across 8 different compilation environments, while improving accuracy by
    \improve over state-of-the-art for differing compilation environments.
    Additionally, when compared with state-of-the-art, \sysname is
    \obfusimprove more accurate when identifying functions in purposefully
    obfuscated binaries.
    Finally, we demonstrate the generality of IOVecs by achieving native
    accuracy when analyzing \texttt{AArch64} binaries using unmodified
    \texttt{x64} IOVecs.

    This paper provides the following contributions:

    \begin{enumerate}
        \item Design of \framework, a framework for semantic binary
        analysis that infers function semantics through program state changes;
        \item \sysname, a practical implementation of \framework that leverages
        coverage-guided, mutational greybox fuzzing to automatically infer
        program states and input structure layouts for functions;
        \item We show the effectiveness of \sysname through a thorough
        evaluation on \coreutils, obfuscated binaries, and cross-architecture
        binaries.
        We further illustrate the effectiveness of \sysname by demonstrating
        that it scales to three large, ubiquitous shared libraries
        in Ubuntu.
    \end{enumerate}


    \section{Challenges and Assumptions} \label{sec:background}

    Here, we outline challenges for semantic function identification,
    and our assumptions when designing \sysname and \framework.

    \subsection{Semantic Function Analysis} \label{ssec:semanticanalysis}

    Reverse engineering a binary can be a long and difficult task.
    The manual process of reverse engineering a binary starts with extracting and
    disassembling the binary instructions, followed by function identification,
    i.e., determining the location and size of functions.
    These tasks can be non-trivial to perform, but recent work~\cite{renovo,
    polyunpack, func_id, func_interface, ida, radare} make them feasible.
    Once the function bounds are identified in the code, the challenging task of
    semantic identification --- determining what computation or program state change
    a function performs --- begins.
    Semantic identification is the hardest, most time-consuming part of reverse
    engineering.


    \begin{figure*}[tb!]
        \centering
        \begin{minipage}[t]{0.55\textwidth}
\begin{lstlisting}[language=C,firstline=10,lastline=18,
            basicstyle=\scriptsize\ttfamily,label={lst:strlensrc},caption={Source}]
size_t strlen(const char *s)
{
	const char *a = s;
	const size_t *w;
	for (; (uintptr_t)s % ALIGN; s++) if (!*s) return s-a;
	for (w = (const void *)s; !HASZERO(*w); w++);
	for (s = (const void *)w; *s; s++);
	return s-a;
}
\end{lstlisting}
        \end{minipage}
        \hspace{1cm}
        \begin{minipage}[t]{0.3\textwidth}
\begin{lstlisting}[language={[x64]Assembler},firstline=7,basicstyle=\scriptsize\ttfamily,
            lastline=17,label={lst:clangO3},caption={\texttt{clang-4.0 -O3}}]
strlen:
	movq	%rdi, %rax
	testb	$7, %dil
	je	.LBB0_4
	movq	%rdi, %rax
	.p2align	4, 0x90
.LBB0_2:
	cmpb	$0, (%rax)
	je	.LBB0_8
	addq	$1, %rax
	testb	$7, %al
\end{lstlisting}
        \end{minipage} \hspace{1cm}
        \begin{minipage}[t]{0.28\textwidth}
\begin{lstlisting}[language={[x64]Assembler}, firstline=7,basicstyle=\scriptsize\ttfamily,
            lastline=17,label={lst:clangO0},caption={\texttt{clang-4.0 -O0}}]
strlen:
	pushq	%rbp
	movq	%rsp, %rbp
	movq	%rdi, -16(%rbp)
	movq	-16(%rbp), %rdi
	movq	%rdi, -24(%rbp)
.LBB0_1:
	movq	-16(%rbp), %rax
	andq	$7, %rax
	cmpq	$0, %rax
	je	.LBB0_6
\end{lstlisting}
        \end{minipage}%
        \hfill
        \begin{minipage}[t]{0.3\textwidth}
\begin{lstlisting}[language={[x64]Assembler}, firstline=7,basicstyle=\scriptsize\ttfamily,
            lastline=17,caption={\texttt{gcc-7.3.0 -O3}},label={lst:gccO3}]
strlen:
	testb	$7, %dil
	movq	%rdi, %r8
	je	.L2
	cmpb	$0, (%rdi)
	jne	.L4
	jmp	.L22
	.p2align 4,,10
	.p2align 3
.L6:
	cmpb	$0, (%rdi)
\end{lstlisting}
        \end{minipage}
        \hspace{1cm}
        \begin{minipage}[t]{0.3\textwidth}
\begin{lstlisting}[language={[x64]Assembler}, firstline=7,basicstyle=\scriptsize\ttfamily,
            lastline=17,caption={\texttt{clang-3.9 -O3}},
            label={lst:clang39}]
strlen:
	testb	$7, %dil
	movq	%rdi, %rax
	je	.LBB0_4
	movq	%rdi, %rax
	.p2align	4, 0x90
.LBB0_2:
	cmpb	$0, (%rax)
	je	.LBB0_8
	incq	%rax
	testb	$7, %al
\end{lstlisting}
        \end{minipage}
        \caption{\texttt{strlen} differences under varying compilation environments.}
        \label{fig:codediffs}
    \end{figure*}

    The largest impediment to semantically recognizing known functions is the
    large code diversity due to different compilation environments.
    Here, we refer to the compilation environment as the exact compiler and linker
    brand and version, optimization level, compile- and link-time flags, linker
    scripts, underlying source, and libraries used to generate a binary.
    Compilers attempt to create efficient, optimized code, and different
    compilers utilize different optimization sets.
    While compilers preserve the high level semantics expressed at the source
    level, the generated binary code is highly variable, as illustrated in
    \autoref{fig:codediffs}.
    The source (\autoref{lst:strlensrc}) is from the \texttt{strlen}
    implementation in the \texttt{musl} C library~\cite{musl}.
    Different optimization levels of the same compiler (\autoref{lst:clangO3} and
    \autoref{lst:clangO0}), different compilers at the same optimization level
    (\autoref{lst:clangO3} and \autoref{lst:gccO3}), and different versions of the
    same compiler and same optimization level (\autoref{lst:clangO3} and
    \autoref{lst:clang39}) all produce different code and/or Control-Flow Graphs (CFGs) from
    the same underlying source.
    Optimizations, like dead code analysis and tail call insertions, also greatly
    affect the generated machine code.
    Even worse, custom function implementations (as opposed to the
    use of system-distributed libraries) will likely produce significantly
    different binaries.


    Compilation environment differences inadvertently conspire to prevent prior
    analyses from trivially transferring to new binaries.
    This forces the analyst to examine every function individually, even for
    functions that were previously analyzed.
    Diversity in compilation environments also presents a significant challenge to
    automated code-based semantic identifiers.
    Static analysis tools, like \texttt{BinDiff} or IDA, are inaccurate when
    presented with highly optimized code~\cite{blanket}.
    State-of-the-art dynamic analysis tools~\cite{blanket, in_memory_fuzzer}
    perform better, but still struggle with varying environments.

    However, regardless of compilation environment, the program state changes a
    function performs \emph{must} remain stable for a binary to exhibit correct
    behavior.
    Barring any bug in the compiler implementation or inconsequential actions
    such as dead stores, the same source code should produce the same
    \textit{semantic} behavior in the final application.
    If this was not the case, binaries would exhibit different, and likely
    incorrect, behavior in different builds.

    Each function implicitly specifies a unique set of input states that produce a
    measurable set of changes in program state, determining the behavior of the
    function.
    Different input states for the same function can produce different changes in
    program state, and different functions may produce different changes in program
    state for a single input state.
    Semantic identification can therefore be defined as the discovery of a
    characteristic set of input states and the corresponding external program state
    changes the function makes given the input.
    We call these inputs and program state changes Input/Output Vectors, or
    \emph{IOVecs}.
    The challenge is to find a sufficiently distinct set of IOVecs that can
    uniquely identify a function.
    However, once found, this set can be used to identify a function in another
    binary regardless of compilation environment.
    IOVecs are thus ideal for semantic function identification.

    \subsection{Assumptions} \label{ssec:assumptions}
    In line with existing semantic analysis tools, when designing \framework, we
    assumed the following:

    \begin{enumerate}
        \item Binary code is stripped, but not packed.
        \item Binary code is generated from a high-level language with functions,
        and function boundaries are known.
        \item Functions make state changes that are externally visible.
        \item The binary follows a discernible and consistent Application
        Boundary Interface (ABI).
        \item Functions do not rely on undefined behavior.
    \end{enumerate}

    When analyzing binaries, reverse engineers start with a likely stripped binary
    from which they infer its behavior.
    The analysts have no access to the underlying source, debugging information,
    symbol table, or any other human-identifiable information.
    We assume the same setting for \framework.
    Semantic analysis frameworks also make the assumption that all code is
    unpacked, and that the binary was generated from a high-level language with
    a notion of individual functions and a known ABI.
    The latter assumption precludes applications written wholly in assembly with no
    discernible functions, and, while packed code is another serious challenge in
    binary analysis~\cite{renovo, polyunpack, ether}, that topic is orthogonal to
    the analysis that semantic analysis frameworks perform.
    Finally, as it is rare in practice and most likely a bug, no code should rely on
    undefined behavior to correctly function.
    The compiler is free to \emph{use} undefined behavior for optimization
    purposes, but the original source should not rely on any specific
    compiler-based optimization utilizing undefined behavior for proper
    functionality.
    Note that functions which rely on randomness (e.g., cryptographic functions)
    are still valid;
    semantic analysis frameworks simply assume that function semantics do not change with the
    compiler.

    Calling a function can cause many writes to occur, however, the writes a
    function makes cannot only be ephemeral, e.g., operating only on local stack
    frame memory and not returning a value.
    This is because once finished, any change would be overwritten or unused by
    later instructions, and thus the program would have been more efficient had it
    not called the function at all.
    Functions that make only ephemeral changes are dead code, and we assume that
    the compiler will simply remove such code from the final binary.
    Depending on optimization level, some program state operations, e.g.,
    dead stores which write to addresses but are never read, can be removed
    from the final binary.
    We do not include such operations in the function's set of program state
    changes, but focus on \textit{persistent} and \textit{externally measurable}
    program state changes.
    We argue that most user space functions conform to these standards, however, we
    discuss the limitations these standards impose in \autoref{sec:disc}.


    \section{\framework} \label{sec:core}

    \framework is a function semantic identification framework, which infers program
    semantics by measuring the effects of execution. Instead of measuring code
    properties, it measures program state changes that result from executing a
    function with a specific initial program state.

    When a function executes, it does so with registers set to specific values, and
    an address space in a particular state, with virtual addresses mapped or
    unmapped to the process' address space, and mapped addresses holding concrete
    values. We refer to the immediate register values and address space state as
    the program state. Every executed instruction does so relative to the current
    program state, and different paths in a function are taken depending on the
    initial program state.

    \framework performs its analysis by instantiating a specific program state
    before function execution, and then measures the program state
    post-execution.
    Measurable program state changes are writes to locations pointed to by pointers,
    data structures, and variables whose valid lifetimes do not end when the
    function returns.
    These types of changes necessarily must be made to registers or memory
    addresses outside the function's stack frame.
    We also consider the immediate return value of a function to be a
    measurable program state change, but exclude changes to general purpose
    registers (e.g., \texttt{rbx} on \texttt{x64}) and state registers (e.g.,
    \texttt{rsp}).
    They are excluded because, for caller-saved general purpose registers,
    their values are immediately irrelevant upon function return, and state
    registers have no bearing on function semantics.
    Additionally, measurable program state changes preclude modifications to
    kernel state not reported to user space.

    While executing, functions make changes to program state as directed by their
    instructions and the current program state.
    A valid program state for one function might cause another function to fault,
    and the same function can perform arbitrarily different actions based on the
    program state upon invocation.
    Therefore, a function implicitly defines the input program states it
    \emph{accepts} --- states where the function can run and return without
    triggering a fatal fault --- and the corresponding output program states based
    upon these input states.
    A function accordingly defines input program states it \emph{rejects} by
    triggering a fault when provided a semantically invalid input state.
    We call these accepting input and corresponding output program states
    Input/Output Vectors, or \emph{IOVecs}.
    A function $A$ is said to accept an IOVec $I$ if $A$ accepts the input program
    state from $I$, and the resulting state from executing $A$ matches the expected
    program state from $I$.
    If either of these conditions do not hold, then $A$ rejects $I$.
    See \autoref{ssec:matching} for the discussion of matching program states.

    Assuming functions make changes to input program states which are measurable
    post-execution, we can reframe semantic function identification.
    Precisely identifying a function can be seen as identifying the full set of
    IOVecs which a function accepts.
    We call that set the \textit{characteristic IOVec set} ($CIS$).

    \begin{figure}[tb]
        \begin{lstlisting}[language=C,label={lst:mystrcpy},caption={An IOVec Motivating
        Example.}]
int my_div(int a, int b, int* c)
{       *c = a / b; return 0;           }
        \end{lstlisting}
    \end{figure}

    Consider the toy example in \autoref{lst:mystrcpy}. An accepting input program
    state is one that has the first argument set to any integer, the second argument
    set to any integer except $0$, and the third argument set to any properly mapped
    memory address. The memory location pointed to by \texttt{c} can initially have
    any value. The corresponding output program state has the return value set to
    $0$, and the memory location pointed to by \texttt{c} contains the value of
    \texttt{a/b}. An IOVec is a single concrete tuple of accepting input state and
    corresponding output state, and $CIS_{\mathtt{my\_div}}$ is the full set of
    IOVecs \texttt{my\_div} accepts.
    Note that only the first two arguments, the
    location pointed to by \texttt{c}, and the return value, are relevant, and that
    neither the full address space nor every register value are relevant.

    Every function has a $CIS$, and we hypothesize that most functions have a unique
    (non-empty) $CIS$. A set of functions that share a $CIS$ is called an
    \textit{equivalence class}. An example of an equivalence class is the set of
    various architecture specific implementations of \texttt{memmove} in
    \texttt{glibc}, e.g., \texttt{\_\_memmove\_sse2} or
    \texttt{\_\_memmove\_avx512}.
    For the sake of brevity, unless otherwise noted, when we refer to a function, we are
    actually referring to an equivalence class of functions with equal
    functionality.

    In the general case, a function's $CIS$ is unbounded. So for practical reasons,
    we attempt to find a subset of a function's $CIS$, which we call the
    \textit{distinguishing characteristic IOVec set}, or $DCIS$. A $DCIS$ for
    function $f$, $DCIS_f$, consists entirely of IOVecs which $f$ accepts, and only
    $f$ accepts every member of $DCIS_f$. Another function, $g$, might accept a
    member of $DCIS_f$, but there is at least one IOVec $I \in DCIS_f$ which $g$
    does not accept.  \framework is used to identify a function \texttt{foo} in a
    binary by providing \texttt{foo} with IOVecs $I_j \in DCIS_f$. If \texttt{foo}
    accepts \textit{all} $I_j$s, then we say that $\mathtt{foo} \equiv f$.

    \framework needs an oracle to provide IOVecs in order to semantically identify
    functions, but there is no definitive source of IOVecs.  \sysname was designed
    to be one such oracle, but other oracles can be devised. For example, IOVecs
    can be derived from unit tests or inferred from a specification. IOVecs also do
    not need to capture large portions of the address space in order for \framework
    to be useful. Instead, IOVecs only need to contain the data that functions
    access, and oracles providing IOVecs can make any attempt at minimizing the data
    in an IOVec.

    The number of IOVecs \framework needs in order to be precise is highly
    dependent on the diversity and number of functions analyzed.
    The minimal theoretical number of needed IOVecs is equal to the number of
    functions being analyzed, because \framework needs at least one accepting
    IOVec to identify and distinguish a function.
    However, it is likely more IOVecs are needed to precisely distinguish
    functions, but, the use of \emph{differences} in semantic behavior for
    discrimination minimizes the number of required IOVecs.

    \subsection{\framework Design} \label{ssec:sedesign}

    \framework performs its analysis in two phases: a coalescing phase and an
    identification phase.
    The coalescing phase, which only needs to be run once, is where functions are
    classified by IOVec acceptances and rejections, and ordered into a binary tree
    accordingly.
    The second phase is where unknown functions are semantically identified by
    providing the unknown functions with specific IOVecs from the binary tree, and
    traversing the tree according to IOVec acceptance.

    \paragraph*{Coalescing Phase}
    \framework starts its analysis by providing every function in its training set
    with every IOVec the oracle provides.
    This establishes a full ground truth of which IOVecs are accepted and
    rejected, ensuring that proper ordering can be achieved.
    When an IOVec is given to a function $f$, one of four results can occur:

    \begin{enumerate}
        \item The function receives a fatal signal (e.g., \texttt{SIGSEGV}), due
        to an improper input program state.
        \item The function does not return before a specified timeout.
        \item The function returns, but the final program state differs from the
        expected output program state.
        \item The function returns, and the final program matches the expected
        output program state.
    \end{enumerate}

    IOVecs that satisfy the last result are added to $DCIS_f$.
    As future work, we want to incorporate rejected IOVecs into the identification
    process, as rejected IOVecs classify the \emph{rejected} semantics of this
    function.

    The result of the coalescing is a proposed $DCIS$ for every function in
    the training set, which is then fed to a decision tree generator.
    The output decision tree contains IOVecs as interior nodes, and functions at
    leaves, and can be used for semantically identifying any number of functions
    later.
    As the tree is generated using \emph{differences} in semantic behavior, it only
    grows linearly in the worst case.
    Every path from root to leaf encodes a minimal $DCIS$ needed to distinguish
    one function from every other in the tree.
    If the same path in the decision tree maps to more than one function, then a
    potential equivalence class exists in the binary.
    The functions in the leaf are those for which the generated $DCIS$ is
    insufficient to fully distinguish one function from another.
    This can be because the generated IOVecs cover the functionality poorly,
    or the functions are truly an equivalence class.

    \paragraph*{Identification Phase}
    To semantically identify functions, the analyst provides the \framework
    implementation with an unknown binary and the generated decision tree from the
    coalescing phase.
    For every function in the unknown binary the following procedure is performed.
    Starting from the root of the decision tree, the IOVec is given to the unknown
    function.
    If the IOVec is accepted, the \textit{true} branch in
    the decision tree is taken;
    otherwise, the \textit{false} branch is taken.
    The unknown function is then tested against another IOVec depending on the
    path taken.
    When the path arrives at a leaf, the unknown function is tested against one more
    IOVec from the leaf function's $DCIS$ for confirmation.
    Again, if the IOVec is accepted, then the function is given the label of the
    function at the leaf.
    If the unknown function gets to a leaf and remains unconfirmed, then the
    function is labeled as unknown.



    \section{\sysname Implementation} \label{sec:timdesign}

    \begin{figure*}[tb!]
        \centering
        \includegraphics[width=\textwidth,height=4.5cm,
        keepaspectratio]{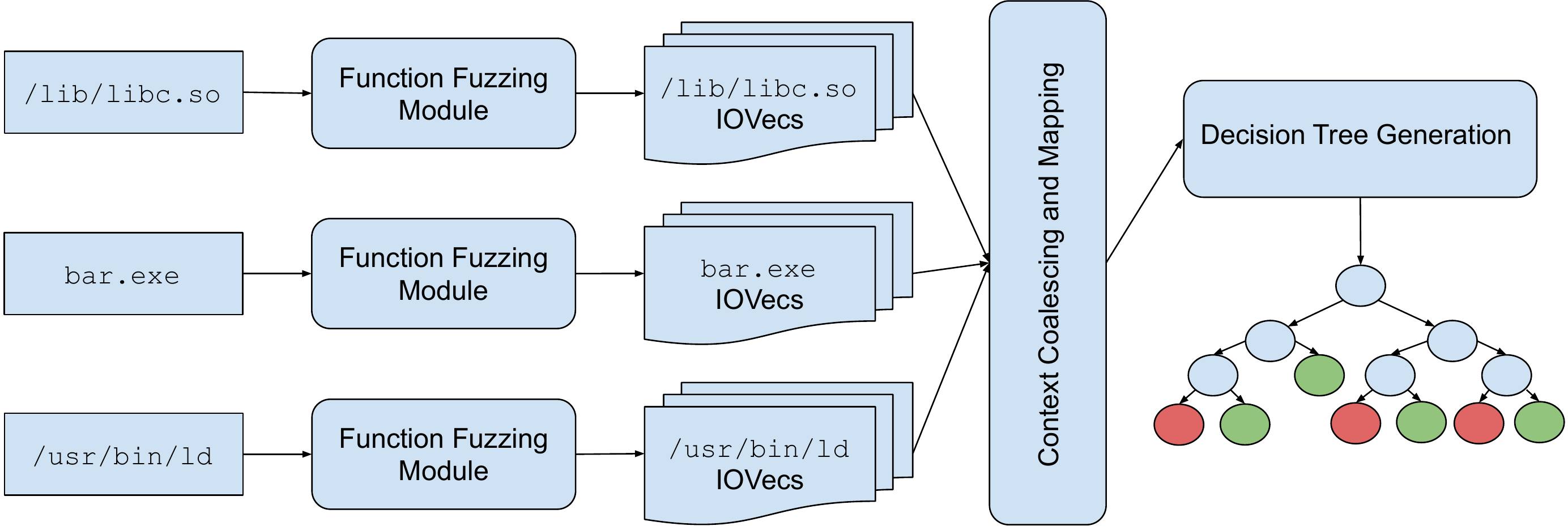}
        \caption{\sysname One-Time Learning Phase.
        The objects in the dotted box are saved for the second phase.}
        \label{fig:phase1}
    \end{figure*}

    \begin{figure*}[bt!]
        \centering
        \includegraphics[width=\textwidth,height=4.5cm,
        keepaspectratio]{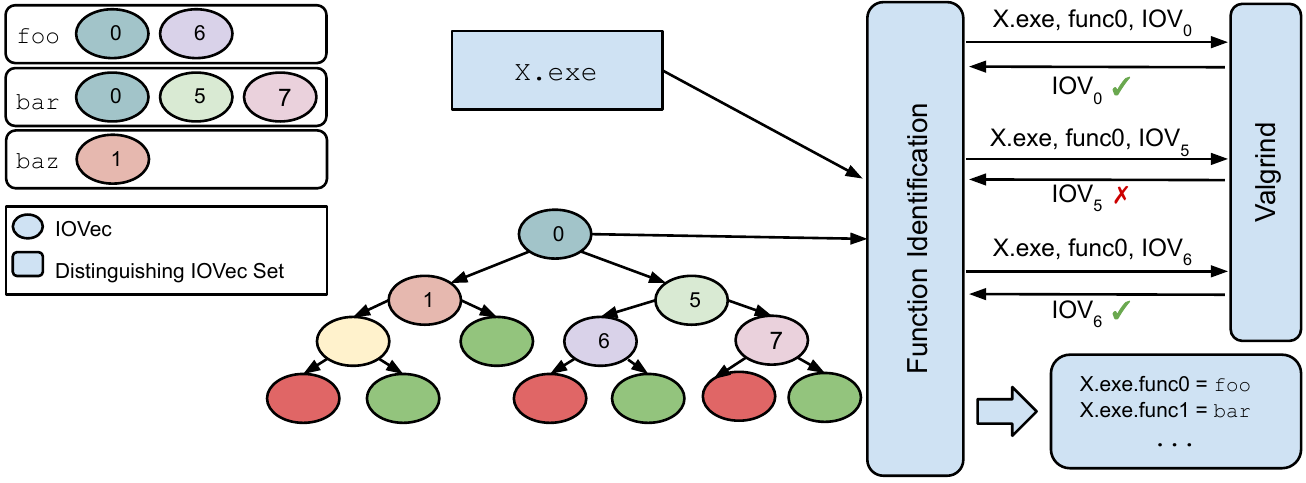}
        \caption{\sysname Identification Phase.
        The \ding{51} and \ding{55} indicates that the IOVec was accepted and
        rejected respectively.
        Paths in the tree leading to green leaves indicate semantic equivalency in the
        unknown binary \texttt{X.exe} to a previously analyzed function
            (\texttt{foo}, \texttt{bar}, or \texttt{baz}), while paths leading to red leaves
            represent unseen/new behavior.}
        \label{fig:phase2}
    \end{figure*}

    \sysname serves as an oracle for \framework.
    Ideally, a formal specification of a function's semantics would allow us to
    generate a minimal $DCIS$.
    However, such a specification is often not available, and we are forced to
    infer valid IOVecs.
    To that end, \sysname uses dynamic instrumentation to precisely control and
    monitor execution, and adapts techniques from state-of-the-art fuzzers to
    generate IOVecs, leveraging code coverage as an optimization and approximation
    of IOVec state coverage.
    \sysname utilizes the knowledge about language concepts, such
    as the calling convention or the existence of pointers.
    This is purely an implementation aspect, and different language semantics would
    have to be accounted for to use \framework.
    For example, the structure of IOVecs will change when analyzing Java
    applications, because there is no concept of a pointer in Java.
    Nevertheless, program state change as a semantic identifier can be applied
    universally.

    With the release of American Fuzzy Lop~\cite{afl} in 2015, fuzzing became an
    invaluable part of many large profile projects~\cite{oss-fuzz}.
    Fuzz testers, or fuzzers, come in many flavors, but mutational coverage-guided
    greybox fuzzers --- the most widely used class of fuzzers
    today ~\cite{afl,honggfuzz,kafl,fs_fuzzing, quickfuzz,
    whitebox_fuzzer,difuze,song2019periscope,tfuzz,angora,redqueen,libFuzzer,
    compiler_fuzzer}
    ---
    are the most relevant to \sysname.
    Mutational fuzzers do not require any specification or expert knowledge of the
    target application;
    a small input file containing a few bytes often generates high code path
    coverage.
    Since we have no information about an unknown function's semantic behavior, the
    ideas behind feedback-guided mutational fuzzing are useful in discovering
    IOVecs.
    By rapidly feeding a function random inputs, and measuring the program state
    change post-execution, we can build a corpus of function identification data
    without any \textit{a priori} knowledge.
    We chose fuzzing as our exploration strategy because fuzzing is optimized to
    maximize code coverage, leading to maximal program state change coverage.
    We do not need full path or code coverage to be accurate, only enough program
    state change coverage (i.e., data coverage) to differentiate semantics.

    \subsection{Exploration Phase} \label{sec:learning}
    \autoref{fig:phase1} shows the overall design of the first phase of \sysname's
    binary analysis.
    \sysname supports analyzing any executable code, including shared libraries.
    Static libraries need to be included in either a shared library or executable.
    \sysname requires neither the source nor any debug information, however,
    it does need boundary information of each function in an executable, or
    the exported symbol names in a shared library. Recent work shows that this
    information can be recovered even for stripped
    binaries~\cite{func_id,func_interface}.
    Once provided with this information, \sysname
    enters its exploration phase, employing coverage-guided fuzzing as a
    way to infer valid IOVecs.

    For each Function Under Test (FUT), \sysname fuzzes the input arguments
    and non-pointer memory object data if any have been deduced, and then
    begins executing the FUT with this randomized program state.
    If that program state is accepted, then the newly discovered IOVec is
    returned, and \sysname examines the coverage it produced.
    If the IOVec produced new coverage, it is added to the FUT's $DCIS$,
    otherwise, it is discarded.
    Either way, the IOVec in the FUT's $DCIS$ that produced the most coverage
    (or a completely new, randomized IOVec in case the $DCIS$ is empty)
    is chosen as a seed for additional fuzzing.
    This process continues until the code coverage exceeds a user-defined
    threshold.
    \sysname supports arbitrary fuzzers that allow intermediate seed export, and
    we similarly support different coverage metrics of the underlying fuzzer.

    \begin{figure}[!tb]
        \footnotesize
        \centering
        \begin{tabular}{|l|l|}
            \multicolumn{1}{c}{\textbf{IOVec Data}} &
            \multicolumn{1}{c}{\textbf{Use}} \\ \hline
            Random seed & Input program state initialization \\
            \rowcolor[HTML]{ECF4FF}
            Pointer input arguments & Input program state initialization \\
            Memory object information & Input program state initialization \\
            \rowcolor[HTML]{ECF4FF}
            Code coverage & Fuzzer seed selection \\
            Expected return value & Post-execution state comparison \\
            \rowcolor[HTML]{ECF4FF}
            Expected memory state byte values & Post-execution state
            comparison \\
            Unique system calls & Post-execution state comparison \\
            \rowcolor[HTML]{ECF4FF}
            Originating architecture & IOVec translation \\
            \hline
        \end{tabular}
        \caption{Data stored in IOVecs.}
        \label{fig:ioveccomp}
    \end{figure}

    \sysname stores the input program state and expected program state
    post-execution in an IOVec.
    Storing the entire address space is both a waste of storage and imprecise.
    Instead, IOVecs save the data listed in \autoref{fig:ioveccomp}.
    Memory object information is the coarse-grained input layout and
    global memory objects inferred during the generation of the IOVec, and
    includes location, size, and pointer sub-member offsets.
    While generating IOVecs, \sysname uses code coverage to select an IOVec
    to mutate, so we include the instructions executed by the FUT when
    provided with the IOVec.

    Prior to execution, the program state is initialized according to
    the current IOVec.
    The memory object information for both input and global data is inferred
    during the exploration phase, and is allocated only for the current IOVec.
    Non-pointer values for data structures and function input arguments are
    setup using the random seed stored with the IOVec, while assuring
    that the same initial values are consistently applied across executions.
    Data pointer sub-members are properly assigned based on their locations
    specified by the IOVec, and then the function is executed with the newly
    established memory state.
    While our experiments did not require us to pass arguments into the FUT on
    the stack, we still support such functionality as we have full
    control over the initial stack pointer.
    Once complete, the resulting program state is either stored if generating
    new IOVecs, or compared with the expected program state.

    \subsection{Matching Program States} \label{ssec:matching}
    \begin{figure*}
        \begin{minipage}{.49\textwidth}
            \centering
            \setlength{\tabcolsep}{5pt}
            \begin{tabularx}{\textwidth}{|c|c|c|c|c|}
                \hline
                \multicolumn{1}{l}{Policy} & \multicolumn{1}{l}{Instruction} &
                \multicolumn{1}{l}{$t$ Tainted?} & \multicolumn{1}{l}{$u$
                    Tainted?} &
                \multicolumn{1}{l}{Taint Policy} \\ \hline
                1 & $t = u$ & Yes & No & T($u$); R($t$)  \\
                \rowcolor[HTML]{ECF4FF}
                2 & $t = u$ & No & Yes & \\
                3 & $t = u$ & Yes & Yes & \\
                \rowcolor[HTML]{ECF4FF}
                4 & $t = t \circ u$ & Any & Any & \\ \hline
            \end{tabularx}
            \caption{Backwards Taint Propagation. $t$ and $u$ can be a register or
            memory address. T($x$) taints $x$ and R($x$) removes taint from $x$.
                $\circ$ denotes any logic or arithmetic operator.}
            \label{tbl:taint}
        \end{minipage}
        \hspace{.5cm}
        \vspace{-.1cm}
        \begin{minipage}{.49\textwidth}
            \centering
            \hspace{1cm}
            \vspace{-.25cm}
            \includegraphics[trim=0 0 0 .25cm, height=3cm]{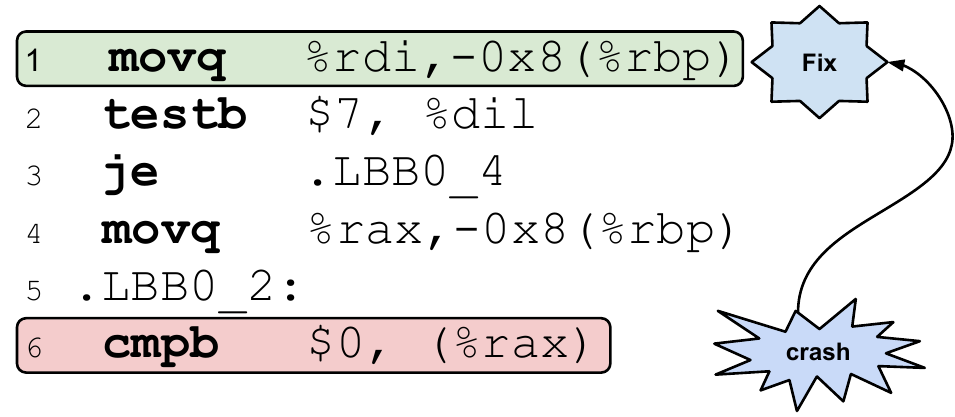}
            \caption{Backwards taint analysis to fix pointer arguments.}
            \label{fig:ptrfix}
        \end{minipage}
    \end{figure*}

    \framework uses matching program states to differentiate and classify
    functions' semantics.
    Here, we present our definition of matching states that \sysname uses
    to identify C functions.

    Recall that our notion of input program state includes memory objects for
    both global data as well as input arguments.
    Semantically similar functions modify memory objects in similar ways (if
    at all), so we capture the resulting memory state of allocated objects
    post-execution.
    Due to our fine-grained control over the memory state, any pointer value
    (either as an input argument or as a structure sub-member) is the same
    across executions.
    The allocated memory objects can be any arbitrary data structure,
    containing a mix of pointer and non-pointer data at various locations
    within the structure.
    Program states match when non-pointer values in memory regions are
    byte-wise the same, and any pointers to sub-objects are located at the same
    offset from the object start.
    If there is a single mismatch in memory objects between two program states,
    then the states do not match.

    Return values are also pertinent, but can be implementation dependent.
    We recognize two types of return values: pointers and non-pointers.
    Due to the lack of any type information in binaries, precisely determining if
    a return value is a pointer is challenging.
    We conservatively test if the return value maps to a readable region in memory,
    and if it does, we designate the return value as a pointer.
    If a return value is not readable in memory, then we consider it a non-pointer,
    and can represent functions that perform raw computations (e.g.,
    \texttt{sin} or \texttt{toupper}), or adhere to a contract (e.g.,
    \texttt{strcmp} which can return any value $< 0$, $= 0$, or $> 0$).

    Finally, because system calls provide services that cannot be provided by
    user-space code and cannot be optimized out, semantically equivalent functions
    \emph{must} invoke the same set of system calls.
    Order and number of system calls made, however, can differ among semantically
    equivalent functions (e.g., calling \texttt{read(fd,~1)} 4 times could be the
    same as calling \texttt{read(fd,~4)} once).
    Therefore, we include the set of unique system calls invoked while executing
    with the specific input as part of the IOVec.
    Semantically equivalent functions must invoke the same set of system calls,
    and can execute neither more nor fewer unique system calls.

    For two program states to match, the values contained in return
    registers must match in the following ways.
    Return values must both be pointers or non-pointers.
    As we do not know the size of the underlying memory region, we do not check the
    underlying memory values if the return values are pointers;
    we simply say the return values match.
    Without more sophisticated analysis, this can be a source of inaccuracy.
    If the return values are non-pointers, they must be equal, or both must be
    positive or negative.
    If all input pointers (including pointers to all sub-objects) match, the
    return values match, and the same set of system calls are invoked, then the two
    program states match.
    As we do not perform any static analysis, \texttt{void} functions will also
    go through return value analysis, leading to another source of potential
    imprecision.

%
%

    \subsection{Pointer Derivation} \label{ssec:ptr}

    A major challenge to generating high-quality IOVecs is the detection of pointers as input.
    As binaries contain no type information, determining if an input
    argument is a pointer is an ongoing research topic~\cite{runtime_check,
    valgrind}.
    Without recovering which arguments are pointers, determining a $DCIS$ for a
    function is generally impossible, and only incomplete behavior will be captured.

    A simple solution would replace an invalid address with a valid address before
    an illegal dereference occurs.
    While such a solution has been successfully used to solve other problems in
    binary analysis~\cite{peng2014x,johnson2013forced}, it would not work in
    \framework, because the underlying problem --- semantically, an input is
    supposed to be a pointer when it is not --- remains unsolved.
    \framework relies on capturing program state changes that arise from executing a
    function with a specific input program state.
    By replacing an illegal address \textit{in situ}, the resulting output program
    state does not necessarily arise from actions performed given the initial state,
    and an IOVec with an input state and an unrelated output state would be
    generated.

    Consider the code in \autoref{fig:ptrfix}, which is adapted from the
    \texttt{strlen} assembly in \autoref{fig:codediffs}.
    The first pointer argument (passed in using register \texttt{rdi}) is stored
    on the stack (line 1).
    Later, that address is written to register \texttt{rax} (line 4), and then is
    dereferenced and compared with the null terminator (line 6).
    Our fuzzing strategy is unlikely to supply a valid address as input, and line 6
    will cause a \texttt{SIGSEGV} signal to be issued.

    The simple approach would replace the invalid address in \texttt{rax} with a
    valid address.
    If the function later returns with no other issue, then \sysname would
    register \texttt{strlen} as accepting the input program state with \texttt{rdi}
    set to a random (non-pointer) value.
    This is incorrect, and during the identification phase, an implementation of
    \texttt{strlen} in an unknown binary would \emph{not} accept the input
    program state.
    That \texttt{strlen} implementation would then be marked with an
    incorrect label.

    The solution we propose is a backwards taint analysis inspired by Wang et
    al.~\cite{in_memory_fuzzer}.
    While generating IOVecs in its exploration phase, \sysname records immediate
    register values before every instruction executes, and, if a segmentation
    fault occurs, we get the register containing the faulty address, which is
    the taint source.
    \sysname then uses the saved register values to propagate the taint back to a
    root sink.
    The taint propagation policy is listed in \autoref{tbl:taint}.
    Starting from the last executed instruction, each instruction is parsed in
    reverse order until all instructions are iterated through.
    The root sink is the last tainted register or memory address after all
    instructions are processed.

    After sink discovery, \sysname searches for previously allocated
    memory objects.
    If no object is found near the faulting address, then \sysname
    builds a new memory object by allocating a fixed-size memory region, and
    records the current location and size of the object.
    \sysname uses this information for inferring new bounds and pointer
    sub-members if another segmentation fault occurs after execution
    restarts. Analysts can use the bounds information for more
    sophisticated analysis after decision tree generation.
    Once the object has been created or updated, \sysname writes the pointer
    to the sink, and begins executing the FUT from its beginning using the
    newly adjusted program state.

    The backwards taint analysis restarts with every segmentation fault until
    the FUT successfully returns.
    When the FUT actually completes, \sysname records the correctly initialized
    input program state, the corresponding output program state, and the
    coarse-grained object structure derived from the backwards taint analysis.
    \sysname only tracks which memory areas are supposed to be pointers, and no
    other semantic meaning is given to memory regions containing non-pointer data.
    Further fuzzing iterations maintain the memory object structure, and only the
    non-pointer memory areas are fuzzed.


    \section{Evaluation} \label{sec:eval}
%
    \pgfplotstableread[row sep=\\,col sep=&]{
        size & n \\
        1 & 509 \\
        2 & 129 \\
        3 & 48 \\
        4 & 24 \\
        5 & 17 \\
        6 & 1 \\
        7 & 10 \\
        8 & 4 \\
        9 & 0 \\
        10+ & 8 \\
    } \equivsizes
    \begin{figure*}[!tb]
        \begin{minipage}{.52\textwidth}
            \begingroup
            \setlength{\tabcolsep}{4.5pt}
            \renewcommand{\arraystretch}{1.2}
            \centering
            \begin{tabular}{|cc|cc|cc|cc|cc|}
                \hline
                \multicolumn{2}{|c|}{\multirow{2}{*}{\diagbox{\scriptsize{Suite}}
                {\scriptsize{D-Tree}}}}
                & \multicolumn{2}{c|}{\textbf{O0}} & \multicolumn{2}{c|}{\textbf{O1}} &
                \multicolumn{2}{c|}{\textbf{O2}} & \multicolumn{2}{c|}{\textbf{O3}} \\
                \multicolumn{2}{|c|}{} &
                \multicolumn{1}{c|}{\footnotesize{\texttt{LLVM}}} &
                \footnotesize{\texttt{gcc}} &
                \multicolumn{1}{c|}{\footnotesize{\texttt{LLVM}}} &
                \footnotesize{\texttt{gcc}} &
                \multicolumn{1}{c|}{\footnotesize{\texttt{LLVM}}} &
                \footnotesize{\texttt{gcc}} &
                \multicolumn{1}{c|}{\footnotesize{\texttt{LLVM}}} &
                \footnotesize{\texttt{gcc}} \\ \hline
                \multirow{2}{*}{\textbf{O0}} &
                \texttt{LLVM} & \textbf{.874} & .829 & .728 & .691 & .702 & .667 &
                .694 & .743 \\
                & \cellcolor[HTML]{ECF4FF}\texttt{gcc} & \cellcolor[HTML]{ECF4FF}.852 & \cellcolor[HTML]{ECF4FF}\textbf{.851} & \cellcolor[HTML]{ECF4FF}.726 & \cellcolor[HTML]{ECF4FF}.685 & \cellcolor[HTML]{ECF4FF}.691 & \cellcolor[HTML]{ECF4FF}.655 &
                \cellcolor[HTML]{ECF4FF}.691 & \cellcolor[HTML]{ECF4FF}.736
                \\ \hline
                \multirow{2}{*}{\textbf{O1}}
                & \texttt{LLVM} & .661 & .692 & \textbf{.891} & .636 & .753 & .690 &
                .718 & .671 \\
                & \cellcolor[HTML]{ECF4FF}\texttt{gcc} & \cellcolor[HTML]{ECF4FF}.848 & \cellcolor[HTML]{ECF4FF}.811 & \cellcolor[HTML]{ECF4FF}.815 & \cellcolor[HTML]{ECF4FF}\textbf{.852} & \cellcolor[HTML]{ECF4FF}.808 & \cellcolor[HTML]{ECF4FF}.782 &
                \cellcolor[HTML]{ECF4FF}.804 & \cellcolor[HTML]{ECF4FF}.854
                \\ \hline
                \multirow{2}{*}{\textbf{O2}}
                & \texttt{LLVM} & .723 & .744 & .836 & .736 & \textbf{.929} & .789 &
                .916 & .752 \\
                & \cellcolor[HTML]{ECF4FF}\texttt{gcc} & \cellcolor[HTML]{ECF4FF}.710 & \cellcolor[HTML]{ECF4FF}.757 & \cellcolor[HTML]{ECF4FF}.835 & \cellcolor[HTML]{ECF4FF}.718 & \cellcolor[HTML]{ECF4FF}.828 & \cellcolor[HTML]{ECF4FF}\textbf{.892} &
                \cellcolor[HTML]{ECF4FF}.830 & \cellcolor[HTML]{ECF4FF}.799
                \\ \hline
                \multirow{2}{*}{\textbf{O3}} &
                \texttt{LLVM} & .723 & .742 & .835 & .735 & .929 & .798 &
                \textbf{.926} & .760 \\
                & \cellcolor[HTML]{ECF4FF}\texttt{gcc} & \cellcolor[HTML]{ECF4FF}.849 & \cellcolor[HTML]{ECF4FF}.830 & \cellcolor[HTML]{ECF4FF}.825 & \cellcolor[HTML]{ECF4FF}.819 & \cellcolor[HTML]{ECF4FF}.822 & \cellcolor[HTML]{ECF4FF}.848 &
                \cellcolor[HTML]{ECF4FF}.820 & \cellcolor[HTML]{ECF4FF}\textbf{.932}
                \\ \hline
            \end{tabular}
            \caption{Geometric mean F-Score for \coreutils per decision tree
            compilation environment (rows) across evaluation suite compilation
            environments.
            }
            \label{tab:coreutils}
            \endgroup
        \end{minipage}%
        \hspace{10pt}
        \begin{minipage}{.45\textwidth}
            \centering
            \begin{tikzpicture}
                \begin{axis}[
                    ybar,
                    symbolic x coords={1,2,3,4,5,6,7,8,9,10+},
                    xtick=data,
                    width=\columnwidth,
                    height=2.35in,
                    nodes near coords,
                ]
                    \addplot table[x=size,y=n]{\equivsizes};
                \end{axis}
            \end{tikzpicture}
            \caption{Distribution of all equivalence class sizes across all
            decision trees in the \coreutils evaluation.}
            \label{fig:alldist}
        \end{minipage}
    \end{figure*}

    Our evaluation focuses on 64-bit System-V Linux binaries derived from
    C source code.
    Our implementation uses the Valgrind~\cite{valgrind} binary translator,
    with \valgrindloc lines of C code, and \pyloc lines of Python.
    As Valgrind readily supports Windows (through the Windows Subsystem for
    Linux) and 32-bit systems, adding support for these systems would require
    only minor engineering efforts.
    Valgrind supports diverse architectures, and our implementation so far
    covers \texttt{x64} and \texttt{AArch64}.
    \sysname performs its taint analysis using Valgrind's architecture
    independent intermediate representation (VEX), so supporting additional
    architectures does not require significant modifications.
    Instead, developer effort is limited to providing necessary ABI information
    (e.g., the return value register), and a recompilation.

    We performed our evaluation using an Intel Core i7-6700K CPU, with 32 GB of
    RAM, and running Ubuntu 16.04 LTS.
    We address the following research questions:

    \begin{enumerate}
        \item How accurate and scalable is \framework in identifying functions in
        binaries?
        \item Is \framework truly resilient against compilation environment
        diversity?
        \item Do IOVecs generated by \framework apply to other architectures?
        \item Does \sysname create meaningful equivalence classes?
    \end{enumerate}

    Our results do in fact show that \framework is a feasible and accurate
    semantic function identifier.
    Additionally, our results show that \framework is largely unaffected by
    compilation environment changes, and that \framework can quickly identify
    previously analyzed functions.
    Our large-scale real-world application evaluation shows that \framework
    can scale to large binaries.
    Finally, we show that IOVecs truly preserve semantics by achieving
    high accuracy when identifying functions in both purposefully obfuscated
    and \texttt{AArch64} binaries.

    \subsection{Accuracy Amid Environment Changes} \label{sec:accuracy}

    To conduct our evaluation of \sysname's accuracy, we selected
    \texttt{wc}, \texttt{realpath}, and \texttt{uniq} from \coreutils, which
    represent medium-sized applications using the default
    compilation environment.
    We compiled the set of applications using \gcc~\cite{gcc} and
    \clang~\cite{llvm}, at \texttt{O0}--\texttt{O3} optimization levels.
    We then built a decision tree (see \autoref{sec:learning}) for each
    application, for a total of \dtreecount decision trees.
    The total amount of fuzzing time allocated for generating IOVecs was
    limited to 5 hours, after which the coalescing phase was allowed as much
    time as necessary.
    See \autoref{ssec:time} for timing details.
    Each tree was used to identify functions in \texttt{du}, \texttt{dir},
    \texttt{ls}, \texttt{ptx}, \texttt{sort}, \texttt{true},
    \texttt{logname}, \texttt{whoami}, \texttt{uname}, and \texttt{dirname},
    each also compiled using \gcc and \clang at \texttt{O0}--\texttt{O3}
    optimization levels.
    These applications represent the \evalcount largest and smallest
    applications as determined by the default \texttt{coreutils} compilation
    environment.
    In order to establish ground truth, we compiled all binaries with debug
    symbols enabled.
    However, \sysname does not use them for its analyses, and they were
    only used for determining accuracy after all analyses had completed.

    We report the geometric mean F-Score (harmonic mean of precision and
    recall) across all compilation environments.
    In order to determine the correctness of a label, we performed a simple
    string comparison between the name of the FUT and the functions in the
    assigned equivalence class.
    If any matched, we record the function name as the assigned label,
    otherwise we use the name of the first function in the equivalence class
    as the assigned label.
    If a function is not matched to an equivalence class, we label the
    function as ``\emph{Unknown}''.
    We then search for the function name among all the classified functions
    in the decision tree.
    The ground truth label is the function name if it appears in the
    classified function list, or ``\emph{Unknown}'' if the function name is
    not in the classified list.
    The classification labels and ground truth labels are then given to the
    \texttt{sklearn.metrics} Python module for F-Score calculation.

    \autoref{tab:coreutils} shows the geometric mean F-Score \sysname achieved
    with decision trees from a specific compilation environment.
    Each row reports the accuracy of all decision trees from the specific
    compilation environment has when used to identify functions in binaries
    generated with a specific compilation environment (presented as the columns).
    The diagonal numbers (in bold) are, therefore, the accuracy rates when the
    decision trees and evaluation suite match in both compiler and optimization
    level.
    They are unsurprisingly among the most accurate \sysname achieved, and
    represent the data most reported by related work.
    Overall, we achieve a high \accuracy accuracy rate, but the diagonal numbers
    --- the numbers which allow for the best apples-to-apples comparison with
    related works --- is \diagaccuracy.
    The generally high F-Scores across compilation environments indicate that
    our accuracy largely comes from \sysname's ability to identify functions
    it has classified, and not from simply assigning an unknown classification
    to functions it has not identified.
    These results show that \framework is accurate as a semantic function
    identifier, as well as largely resilient to compilation environments
    (Research Questions 1 and 2).

    Unfortunately, the two closest systems to \sysname, \emph{BLEX}~\cite{blanket}
    and \emph{IMF-SIM}~\cite{in_memory_fuzzer}, are not available publicly.
    The \emph{BLEX} authors supplied us their code, but it required
    significant engineering to execute with currently distributed Python modules.
    We invested two weeks of development and evaluation time.
    The accuracy we evaluated was much lower, but this could be attributed to
    the required engineering changes or changes in the imported modules.
    The \emph{IMF-SIM} authors remained unresponsive.
    We, therefore, base our comparison with related work on the
    published numbers, and call for open-sourcing of research prototypes.
    The \emph{BLEX} authors report an average accuracy of $.50$--$.64$ across
    three compilers (they added Intel's \texttt{icc} compiler) and four
    optimization levels, and the \emph{IMF-SIM} authors report an average
    accuracy of $.57$--$.66$ across three compilers and three optimization
    levels.
    Both systems attempt to build a classification vector from code measurements,
    and their lowest accuracies come from labeling functions in binaries from
    compilation environments different from their source models.
    \sysname, in contrast, is accurate regardless of compilation environment,
    as evidenced by the off-diagonal numbers in \autoref{tab:coreutils}.
    With a geometric mean accuracy of \offdiagaccuracy, our results show an
    average \improve increase in accuracy in differing compilation
    environments over these works.
    The inaccuracy in \emph{BLEX} and \emph{IMF-SIM} arises from the fact that
    code measurements are not a true reflection of function semantics, but are
    instead one way to express function semantics from a large and diverse space
    of possible semantic expressions.
    The trained models they generate become inaccurate when presented differently
    optimized code, because they only capture a small portion of the possible
    semantic expression space.
    \sysname achieves its accuracy by actually measuring a function's semantics
    through program state change, and does not approximate function semantics
    through code measurements.

    Despite its higher accuracy, \sysname does have inaccuracy.
    We identify two major sources of inaccuracy: an overly strict program
    state comparison, and kernel state dependence leading to low-quality IOVecs.

    \paragraph*{Strict Program State Comparison}
    In \autoref{ssec:matching}, we detailed our policy for comparing program
    states, which we use in lieu of code measurements for determining semantic
    similarity.
    We opted for a strict policy where both return values and allocated
    memory areas must match exactly in order for an IOVec to be accepted.
    However, at lower optimization levels, we might capture dead stores that are
    optimized out at higher optimization levels.
    For example, the \texttt{c\_isprint} function, which returns a single byte,
    contains an additional \texttt{movzx} instruction in \texttt{O0} not
    present in any later optimization level.
    This instruction operates on the return register, which changes the higher
    order bits, while higher optimization levels simply write to the lowest byte
    in the return register without changing any further bit value.
    The write to the higher order bits is a dead store, since any caller will
    only ever read the lowest byte of the return register.
    However, we capture this behavior in an IOVec, and our strict return value
    comparison policy often determines the return values to be different, leading
    to a mislabel.
    This is not a fundamental flaw with \framework, but an artifact of our
    program state matching policy.
    A different policy that more precisely compares relevant program state could
    better account for inconsequential program state changes.

    \paragraph*{Kernel State Dependence}
    For simplicity, we designed \sysname to assume nothing when generating IOVecs,
    and it always executes functions in isolation.
    However, there are functions (e.g., \texttt{close} and \texttt{munmap}) that
    depend on the results of previous functions in order for the input arguments
    to be semantically correct.
    For instance, \texttt{close} requires that the input integer be a valid open
    file descriptor (as obtained from \texttt{open}), and any input that is
    \emph{not} a valid file descriptor is semantically incorrect.
    Because \sysname does not perform any initial setup to obtain semantically
    correct input values, any IOVec generated for these functions only exercise
    the error checking functionality, which is likely to be similar to many other
    functions.
    This has two negative effects: unrelated functions get grouped into an
    equivalence class, and unrelated FUTs can be assigned to this equivalence
    class simply because they share similar error handling behavior.
    This is, again, not a fundamental flaw in \framework, but instead is a result of
    \sysname's focus on user-space functions.
    We expect that our accuracy would improve significantly if we added some
    common environmental activities (e.g., opening file descriptors or memory
    mapping address spaces) to our IOVec design.
    We keep it as future work to incorporate application specific environmental
    setup to \sysname.


    \subsection{Equivalence Class Distributions} \label{ssec:dist}

    \setlength{\tabcolsep}{5.5pt}
    \begin{figure}[!tb]
        \begin{tabular}{|c|cc|cc|cc|cc|}
            \hline
            & \multicolumn{2}{c|}{\textbf{O0}} & \multicolumn{2}{c|}{\textbf{O1}}
            & \multicolumn{2}{c|}{\textbf{O2}} & \multicolumn{2}{c|}{\textbf{O3}} \\
            & \multicolumn{1}{c|}{\footnotesize{\texttt{LLVM}}} &
            \footnotesize{\texttt{gcc}} &
            \multicolumn{1}{c|}{\footnotesize{\texttt{LLVM}}} &
            \footnotesize{\texttt{gcc}} &
            \multicolumn{1}{c|}{\footnotesize{\texttt{LLVM}}} &
            \footnotesize{\texttt{gcc}} &
            \multicolumn{1}{c|}{\footnotesize{\texttt{LLVM}}} &
            \footnotesize{\texttt{gcc}} \\ \hline
            \small{$N$} & \small{78} & \small{73} & \small{72} & \small{52} &
            \small{38} & \small{32} & \small{36} & \small{40} \\
            \rowcolor[HTML]{ECF4FF}
            \small{$\overline{N}$} & \small{1.76} & \small{1.85} &
            \small{1.82} & \small{1.68} & \small{1.78} & \small{1.47} &
            \small{1.69} & \small{1.69} \\
            \hline
        \end{tabular}
        \caption{Geometric mean count of classified functions ($N$), average
        number of functions per equivalence class ($\overline{N}$) for all
        \coreutils generated decision trees.
            The median equivalence class size is 1.00 for all decision trees.}
        \label{tab:equivclasses}
    \end{figure}

    \autoref{tab:equivclasses} shows the geometric mean number of classified
    functions ($N$), and the average number of functions per equivalence class
    ($\overline{N}$).
    Ideally, $\overline{N}$ should be close to one, as most functions provide unique
    and singular functionality, and thus should be assigned as the sole member of an
    unique equivalence class.
    However, with the existence of wrapper functions, it is likely $\overline{N}$ will be
    higher.
    It nevertheless should be low, because one could trivially get high accuracy
    by grouping all functions into the same equivalence class.
    As \autoref{tab:equivclasses} shows, we achieve a low $\overline{N}$ across
    our decision trees, which indicates that our fuzzing strategy is a
    generally sound technique for generating sufficiently distinctive IOVecs.
    Additionally, the equivalence class size distributions in
    \autoref{fig:alldist} show that we are creating hundreds of equivalence
    classes with one or two functions per equivalence class, which provides
    evidence that we satisfy Research Question 4.
    We, therefore, claim that our accuracy comes from \sysname's ability to
    distinguish function semantics, and that \sysname does not simply group all
    functions into a few equivalence classes.

    There are equivalence classes containing a large (10+) number of functions.
    These are cases where our fuzzing strategy was unable to trigger deep
    functionality, yet the classified functions share a common failure mode
    (e.g., return $-1$ for invalid input), or very similar functionality.
    For example, there is a 12 function sized equivalence class in the
    \texttt{realpath clang-O1} decision tree that contains 8 functions
    \texttt{strcaseeq[0-7]} that perform the same action with increasingly
    fewer input arguments.
    Improvements in related fuzzing work, especially works that improve deep
    code coverage~\cite{angora,Matryoshka}, will directly translate to an
    improvement of IOVec generation, and a reduction of the size of these
    equivalence classes.

    \subsection{Training and Labeling Time} \label{ssec:time}
    \sysname is scalable in both training time and storage requirements.
    On average, \sysname takes \dtreetime to generate a decision tree, which
    includes generating IOVecs and the coalescing phase described in
    \autoref{sec:learning}.
    As stated before, however, this analysis only needs to be done one time.
    Once the decision tree is generated, semantic analysis is very quick, taking,
    on average, only \labeltime to classify a binary in the evaluation set.
    Additionally, all operations in both of \sysname's phases represent completely
    independent work loads, and as such are embarrassingly parallel.
    Therefore, execution time varies with the available hardware.
    Furthermore, the generated decision tree size is very small, with an
    geometric mean size of \treesize.
    So, while IOVecs have no upper bound in their spatial size as they record the
    memory state of relevant inputs and their sub-members, in practice they are
    small.

    \emph{BLEX} reports $1,368$ CPU hours for training, and 30 CPU minutes to
    classify a binary in \texttt{coreutils}.
    \emph{IMF-SIM} takes $1,027$ CPU hours for training, and 31 CPU minutes to
    classify a \texttt{coreutils} binary.
    Due to significant hardware differences between our respective
    experimental setups, and the lack of available source code for the
    related work, we cannot make any fair quantitative comparison.
    However, we believe that we are faster at semantic queries as we
    organize past analysis in a tree structure;
    \emph{BLEX} and \emph{IMF-SIM} must compare the feature vector they
    record with every past feature vector, creating an $\mathcal{O}(log(n))$ vs.
    $\mathcal{O}(n)$ search performance disparity.
    Neither works report spatial size of their feature vectors, however
    \emph{BLEX} and \emph{IMF-SIM} restrict the number of instructions
    executed, which caps the size of their respective feature vectors.

    \subsection{Equivalence Classes and Imprecision} \label{sec:imprecision}
    \begin{figure}[!tb]
        \footnotesize
        \centering
        \begin{tabular}{|l|l|}
            \hline
            \texttt{quote} & \texttt{quotearg} \\
            \rowcolor[HTML]{ECF4FF}
            \texttt{quotearg\_char} & \texttt{quotearg\_colon} \\
            \texttt{set\_program\_name} & \\
            \hline
        \end{tabular}
        \caption{An equivalence class in an \texttt{LLVM O0} decision tree.}
        \label{fig:casestudy}
    \end{figure}

    To provide a concrete example of an equivalence class, as well as illustrate the
    issues with accuracy, we highlight a specific, large equivalence class
    generated for the \texttt{wc LLVM O0} decision tree.
    The functions comprising this equivalence class are listed in
    \autoref{fig:casestudy}, and include various functions that return
    strings used for printing command line arguments, and a \texttt{void}
    function used to set a global variable containing the name of the called
    application.


    The equivalence class size highlights the effectiveness and limitations of
    our fuzzing strategy and state comparison policy.
    All functions take a \texttt{char*} as their first argument,
    \texttt{quote} and \texttt{quotearg} are both wrappers for the same
    function, \texttt{quotearg\_n\_options}.
    Additionally, \texttt{quotearg\_colon} is just a wrapper function for
    \texttt{quotearg\_char}, but unsurprisingly supplies the colon character
    as input, while \texttt{quotearg\_char} is a wrapper for \texttt{quotearg}.
    \sysname reasonably groups these functions into an equivalence class
    (providing evidence for Research Question 4).
    While our short fuzzing campaigns did not capture the subtle
    differences between the various functions, longer campaigns or better
    fuzzers will likely do so.

    \texttt{set\_program\_name} is placed in this equivalence class
    (despite being a \texttt{void} function) because the return register is
    used as a scratch register holding a valid pointer.
    Therefore, our program state equivalence check detailed in
    \autoref{ssec:matching} passes for all generated IOVecs.
    As future work we plan to incorporate detailed global memory accesses
    into IOVecs, which should prevent such cases.
    \texttt{set\_program\_name} illustrates a common source of
    inaccuracy when using classification data from differing binaries.
    Recall that the decision trees were created using a single binary.
    Even though another binary might have the same function present, \sysname
    will likely mislabel that version of \texttt{set\_program\_name}, because
    the IOVecs generated for that function will encode the name of the source
    binary as part of the expected program state.
    Other binaries will likely not be named the same, and thus our program
    state matching policy detailed in \autoref{ssec:matching} will
    erroneously determine the two functions as different.
    Luckily, as our evaluation shows, these types of functions are rare in
    practice.



    \section{Case Studies} \label{sec:casestudies}

    We provide three case studies that demonstrate the effectiveness of our
    approach.

    \subsection{Accuracy Against Obfuscated Code} \label{sec:obfus}
    Malware authors will often employ code obfuscation to impede binary
    analysis~\cite{banescu2016code,syntia}.
    Code obfuscation attempts to hide semantic meaning through code
    transformations, such as adding unrelated control-flow or instruction
    substitution, while still preserving the intended function semantics.
    Code-based semantic analysis can be stymied when attempting to identify
    purposefully obfuscated code, because the resulting code is far from
    ``normal,'' and thus hard to correlate with models derived from
    unobfuscated binaries.
    \framework, however, relies on \emph{semantic} (rather than code)
    measurements guaranteed to be preserved by code obfuscators.
    Therefore, \sysname should largely be unaffected by code obfuscation.

    To test this hypothesis, we compiled our \texttt{coreutils} suite
    (\texttt{du}, \texttt{dir}, \texttt{ls}, \texttt{ptx}, \texttt{sort},
    \texttt{true}, \texttt{logname}, \texttt{whoami}, \texttt{uname}, and
    \texttt{dirname}) using the LLVM-Obfuscator (OLLVM)~\cite{llvmobfus} at
    \texttt{O2}, enabling separately the bogus control-flow (bcf),
    control-flow flattening (fla), and instruction substitution (sub)
    obfuscations.
    Following the experimental methodology of the \emph{IMF-SIM} authors, we
    used the \texttt{O0} decision trees to measure semantic function
    identification accuracy in each of the three respective obfuscated binaries,
    using the same accuracy measurement metric described in
    \autoref{sec:accuracy}.

    The results are listed in \autoref{fig:obfus}.
    We match or exceed the results achieved by \textit{IMF-SIM}, with an average
    increase in accuracy of \textbf{\obfusimprove}.
    As our accuracy against obfuscated binaries closely matches our accuracy against
    unobfuscated binaries, we provide evidence that \sysname (and, by extension,
    \framework in general) is unaffected by existing obfuscation techniques.
    Any inaccuracy when identifying functions in obfuscated binaries comes from the
    same sources as analyzing normal binaries, as discussed in \autoref{sec:accuracy}
    and \autoref{sec:imprecision}.
    Furthermore, these results also give evidence that Research Question 2 is answered,
    as not only are the binaries purposefully obfuscated, but are also compiled using a
    much older version of LLVM than our evaluation version.

    \begin{figure}[!tb]
        \centering
        \begin{tabular}{|c|c|c|c|c|}
            \hline
            & & \sysname & \textit{IMF-SIM} & \% Difference \\ \hline
            & bcf & 0.787 & 0.385 & 105  \\ \cline{2-5}
            & \cellcolor[HTML]{ECF4FF}fla & \cellcolor[HTML]{ECF4FF}0.772 &
            \cellcolor[HTML]{ECF4FF}0.576 & \cellcolor[HTML]{ECF4FF}34.1 \\
            \multirow{-3}{*}{\rotatebox[origin=c]{90}{\texttt{gcc}}} & sub
            & 0.752 & 0.664 & 13.2 \\
            \hline
            & bcf & 0.806 & 0.513 & 57.1 \\ \cline{2-5}
            & \cellcolor[HTML]{ECF4FF}fla & \cellcolor[HTML]{ECF4FF}0.795 &
            \cellcolor[HTML]{ECF4FF}0.649 & \cellcolor[HTML]{ECF4FF}22.5 \\
            \multirow{-3}{*}{\rotatebox[origin=c]{90}{\texttt{LLVM}}} & sub
            & 0.813 & 0.779 & 4.30 \\
            \hline
        \end{tabular}
        \caption{Obfuscated Code Accuracy Comparison}
        \label{fig:obfus}
    \end{figure}

    \subsection{\texttt{AArch64} Evaluation} \label{sec:aarch}

    \setlength{\tabcolsep}{5.5pt}
    \begin{figure}[!tb]
        \begin{tabular}{|c|cc|cc|cc|cc|}
            \hline
            & \multicolumn{2}{c|}{\textbf{O0}} & \multicolumn{2}{c|}{\textbf{O1}}
            & \multicolumn{2}{c|}{\textbf{O2}} & \multicolumn{2}{c|}{\textbf{O3}} \\
            & \multicolumn{1}{c|}{\footnotesize{\texttt{LLVM}}} &
            \footnotesize{\texttt{gcc}} &
            \multicolumn{1}{c|}{\footnotesize{\texttt{LLVM}}} &
            \footnotesize{\texttt{gcc}} &
            \multicolumn{1}{c|}{\footnotesize{\texttt{LLVM}}} &
            \footnotesize{\texttt{gcc}} &
            \multicolumn{1}{c|}{\footnotesize{\texttt{LLVM}}} &
            \footnotesize{\texttt{gcc}} \\ \hline
            1 & \small{.835} & \small{.805} & \small{.789} &
            \small{.840} & \small{.797} & \small{.803} & \small{.795} &
            \small{.860} \\
            \rowcolor[HTML]{ECF4FF}
            2 & \small{.820} & \small{.803} &
            \small{.766} & \small{.794} & \small{.740} & \small{.761} &
            \small{.737} & \small{.842} \\
            3 & \small{.880} & \small{.866} &
            \small{.833} & \small{.791} & \small{.799} & \small{.849} &
            \small{.796} & \small{.877} \\
            \hline
        \end{tabular}
        \caption{F-Scores for identifying functions in \texttt{coreutils-gcc-O3}
            \texttt{AArch64} binaries using decision trees generated from
            \texttt{x64} \texttt{wc} (1), \texttt{realpath} (2), and
            \texttt{uniq} (3).}
        \label{tab:aarch64}
    \end{figure}

    Function semantics are mainly determined by the high level source code
    written by the programmer, and remain largely constant across architectures.
    Unless explicitly dictated by the programmer through the use of
    preprocessing macros, the same function compiled for one architecture
    will perform the same corresponding program state change in another.
    How the input state is established, and how the resulting program state
    is determined post-execution will change with architecture, but semantics do
    not --- \texttt{x86 strlen}, for instance, does not suddenly change
    functionality when compiled for \texttt{Mips}.
    Therefore, an IOVec generated for one architecture is usable for
    another architecture, as long as there is a suitable IOVec translation
    between the two.

    To that end, we implemented a simple IOVec translation layer between
    \texttt{x64} and \texttt{AArch64}.
    The translation consists solely of a mapping between the argument passing
    registers and return register used by the two architectures, e.g.,
    \texttt{rsi} on \texttt{x64} is mapped to \texttt{x1} in
    \texttt{AArch64}, and \texttt{rax} is mapped to \texttt{x0}.
    As we do not yet support stack argument passing, only the smaller set of
    argument registers are translated, which may slightly lower accuracy.
    However, as most functions have a relatively low arity, our translation
    layer is adequate for the majority of semantic identification tasks.

    We evaluated \sysname's cross-architecture accuracy by compiling the
    \texttt{du} and \texttt{dirname} (the largest and smallest binaries in
    our evaluation suite) on a Raspberry Pi 3 Model B Rev 1.2 running Ubuntu
    20.04 using the ARM \texttt{gcc-9.3.0} compiler at \texttt{O3} optimization.
    We then used the unmodified decision trees generated for the evaluation
    described in~\autoref{sec:accuracy} to identify functions in the ARM
    binaries.
    The results are presented in \autoref{tab:aarch64}, with each column
    listing the accuracy achieved using the \texttt{x64} decision tree
    generated with the enumerated compilation environment.

    \sysname achieves a geometric mean F-Score of \aarchaccuracy across all the
    evaluated binaries, similar to our native geometric mean of \accuracy.
    As our accuracy is largely unaffected by architecture, we strengthen our
    claim that \sysname captures function \emph{semantics}, and provide
    evidence that we answer Research Question 3.
    Additionally, we also provide further evidence that we answer Research
    Question 1, as the \texttt{gcc} version used for this evaluation
    differs from the version used to generate the decision trees.

    \subsection{Large Shared Libraries} \label{sec:libraries}

    \setlength{\tabcolsep}{5.5pt}
    \begin{figure}[!tb]
        \begin{tabular}{|c|cc|cc|cc|cc|}
            \hline
            & \multicolumn{2}{c|}{\textbf{O0}} & \multicolumn{2}{c|}{\textbf{O1}}
            & \multicolumn{2}{c|}{\textbf{O2}} & \multicolumn{2}{c|}{\textbf{O3}} \\
            & \multicolumn{1}{c|}{\footnotesize{\texttt{LLVM}}} &
            \footnotesize{\texttt{gcc}} &
            \multicolumn{1}{c|}{\footnotesize{\texttt{LLVM}}} &
            \footnotesize{\texttt{gcc}} &
            \multicolumn{1}{c|}{\footnotesize{\texttt{LLVM}}} &
            \footnotesize{\texttt{gcc}} &
            \multicolumn{1}{c|}{\footnotesize{\texttt{LLVM}}} &
            \footnotesize{\texttt{gcc}} \\ \hline
            A & \small{-} & \small{.871} & \small{.717} &
            \small{.850} & \small{.759} & \small{.746} & \small{.765} &
            \small{.772} \\
            \rowcolor[HTML]{ECF4FF}
            B & \small{-} & \small{.781} & \small{.633} &
            \small{.695} & \small{.629} & \small{.642} & \small{.629} &
            \small{.639} \\
            C & \small{-} & \small{.794} & \small{.699} &
            \small{.802} & \small{.701} & \small{.722} & \small{.700} &
            \small{.733} \\ \hline
        \end{tabular}
        \caption{F-Scores identifying functions in \texttt{libz} (A),
            \texttt{libpng} (B), and \texttt{libxml2} (C) using a
            \texttt{clang-O0} decision tree.
        We did not evaluate against the \texttt{clang-O0} binary.}
        \label{tab:libraries}
    \end{figure}
    \begin{figure}[!tb]
        \centering
        \begin{tabular}{|c|c|c|c|}
            \hline
            \cellcolor{white} & \texttt{libz} & \texttt{libpng} &
            \texttt{libxml2} \\
            \hline
            $N$ & 126 & 390 & 2080 \\ \rowcolor[HTML]{ECF4FF}
            $\overline{N}$ & 2.47 & 2.48 & 2.44 \\
            $T$ & 17.0 & 25.4 & 158 \\ \hline
        \end{tabular}
        \caption{Decision tree ($N$), average equivalence class sizes
            ($\overline{N}$), and CPU hours needed to generate the decision
            tree ($T$) for large binaries.}
        \label{tab:librariessizes}
    \end{figure}

    An analysis framework that does not scale is unlikely to be useful to
    researchers and engineers.
    Our \texttt{coreutils} evaluation enables a comparison with related work.
    However, we want to demonstrate the scalability of \sysname to larger,
    more complex binaries.

    We chose \texttt{zlib}, \texttt{libpng}, and \texttt{libxml2} as a set of
    shared libraries that are ubiquitous and among the largest distributed
    with Ubuntu.
    We compiled each library using \gcc and \clang at
    \texttt{O0}--\texttt{O3} optimization levels, generated a decision tree
    for the \texttt{clang-O0} binary, and identified functions in the
    remaining binaries.
    Due to the larger size of the binaries involved, we allowed the fuzzing
    campaign to execute for 10 hours, and provided as much time as needed
    for coalescing.
    In order to handle the significant increase in functions, we used a machine
with 45GB memory to generate 
    the decision tree for \texttt{libxml2} (running Debian 9.3 on an Intel
Xeon 3106).
    The machine listed in \autoref{sec:accuracy} was used for all other
    evaluation tasks.
    The 50\% increase in memory to process at least a 10x increase function count
    is a reasonable cost, and does not detract from our scalability claim.

    \autoref{tab:libraries} and \autoref{tab:librariessizes} list the accuracy
    measured (using the same accuracy metric at in \autoref{sec:accuracy}),
    along with the number of functions classified ($N$), average number of
    functions per equivalence class ($\overline{N}$), and CPU time required
    to generate the decision tree ($T$).
    \sysname achieves similar F-Scores as in our \texttt{coreutils}
    evaluation, demonstrating the accuracy and scalability of our approach
    (Research Question 1).
    However, the number of functions per equivalence class is higher than our
    \texttt{coreutils} evaluation.
    This is a consequence of our simplistic coverage-guided fuzzer, as well
    as increased genuine similar functionality.
    For example, there are functions in \texttt{zlib} (e.g.,
    \texttt{gzoffset} and \texttt{gzoffset64}) which only differ in
    the bit count of their input arguments, but otherwise perform the same
    action.
    There are also a large group of functions which first performs a sanity
    check on the input.
    The fuzzer did not create inputs to pass these checks, and the functions are grouped
    into an equivalence class.
    Although inferring valid input is an ongoing research
    topic~\cite{tfuzz,angora,Matryoshka}, both of these problems can be
    mitigated with a longer fuzzing campaign and/or a more sophisticated fuzzer.

    \section{Discussion} \label{sec:disc}

    Here we provide discussion on the limitations of \framework, and on when a function is
    designated as unknown.

    \paragraph*{Limitations}
    We have identified a few sets of functions that \framework is unlikely to
    classify or identify correctly.
    These functions are highly dependent upon the system environment and execution
    context while
    generating IOVecs, as well as during the identification phase.
    Functions like \texttt{getcwd} or \texttt{getuid}, which return the current
    working directory and the user ID respectively, depend on the filesystem,
    current user, and kernel state.
    As these factors differ between runs or are non-deterministic, they violate our
    fundamental assumption --- semantically similar functions change their program
    state in similar ways given a specific input program state.
    To address this limitation, \sysname could model the system state in addition
    to the process state.

    Another set of functions \framework struggles with depend on an initial seed
    being set beforehand.
    Examples of these functions include \texttt{rand} and \texttt{time}.
    As we execute functions without any knowledge about their behavior, we cannot
    provide the seed beforehand as it is difficult to distinguish a seed value
    from other global variables.
    Even if we determine a location of the seed, knowledge of proper
    API usage (e.g., calling \texttt{srand} before \texttt{rand}) is needed to
    correctly use these functions.
    Discerning correct API usage is an active research
    area~\cite{learnapis}, and improvements in this area will directly translate to
    improvements in \sysname.


    \paragraph*{Soundness of \framework}
    When semantic equivalence is determined between two functions, that
    equivalence is only extended as far as the IOVecs tested along the decision
    tree path.
    It is possible that \framework establishes an incorrect semantic equivalence
    between a previously analyzed function \texttt{f}, and a new unseen function
    \texttt{g}, if 1) \texttt{g} accepts all of \texttt{f}'s IOVecs, plus additional
    IOVecs;
    and 2) any additionally accepted IOVec is not along the path to \texttt{f} in
    the decision tree.
    This means that \framework is not a sound technique.
    However, as our equivalence class distributions results show, in practice
    \framework is accurate for most functions, even when functions are similar, as
    with \texttt{strcpy} and \texttt{strncpy}.
    In real world code, most functions have little overlapping functionality,
    which makes \framework a practical tool for semantic identification.
    We have it as future work to incorporate code coverage into the
    semantic similarity analysis, which could produce correct classifications
    through the enforcement of a coverage policy as a condition for semantic
    equivalence.

    \paragraph*{Unknown Functions}
    If a function is encountered that accepts no known $DCIS$, \framework will mark
    this function as unknown.
    When a function is marked as unknown, it can mean one of two things depending
    on the number of accepted IOVecs.
    If the unknown function \emph{never} accepts an IOVec, then it implements wholly
    unknown functionality, and should be a main focus for analysts.
    Otherwise, if the function accepts some IOVecs, then it shares some
    functionality with the functions whose $DCIS$ includes the accepted IOVecs.
    The utility analysts might gain from this information varies with the number
    of IOVecs accepted.
    Many IOVecs rejected with a few IOVec acceptances is likely a common failure
    mode present in many functions, e.g., returning $-1$ on invalid input.
    If many IOVecs in a $DCIS$ are accepted, then the unknown function is likely
    similar to the corresponding function, indicating, e.g., a different version.

    \section{Related Work} \label{sec:rw}

    Similarity analysis is an active area of research~\cite{reoptimization,
    vulseeker,trace_align,binsim,bingo,discovre,eclone,sleuth,graph_bug_search,
    firmup}.
    Jiang et al.~\cite{auto_mining} first proposed using randomized testing in
    function similarity analysis, drawing inspiration from polynomial identity
    testing.
    Their system, which requires source, finds syntactically different yet
    semantically similar code fragments in large (100+ MLOC) code bases.
    \framework does not require source, which, since source is often unavailable when reverse
    engineering binaries, makes \sysname a more practical choice.

    Current state-of-the-art binary analysis tools all rely on code measurements.
    \emph{BLEX}~\cite{blanket} extracts feature vectors of function code, such
    as values read and written to the stack and heap, by guaranteeing that every
    instruction is executed.
    The authors also implemented a search engine with their system similar to
    \sysname.
    Wang, et al.~\cite{in_memory_fuzzer} perform code similarity analysis using a
    system called \emph{IMF-SIM}.
    \emph{IMF-SIM} uses an in-memory fuzzer to measure the same metrics as
    \emph{BLEX}, instead of forcing execution to start at unexecuted instructions.
    As stated in our evaluation, these works still struggle with differing
    compilation environments, while \sysname has consistently high accuracy
    irrespective of compilation environment.
    Both works focus on measuring code properties, which change with different
    compilation environments.
    \framework, in contrast, uses IOVecs, which are independent of code, and
    encodes differing semantics in a binary decision tree.

    Pewny, et al.~\cite{pewny2015cross} compute a signature of a bug,
    and search for that signature in other (possibly different ISA) binaries.
    The signature involves computing inputs and corresponding outputs to basic
    blocks in functions' CFGs through dynamic instrumentation similar to \sysname.
    While the authors admit that semantic function identification is not their
    expected use case, their system can be used as such by supplying a function as
    the ``bug.'' This work, however, heavily relies on the structure of
    the CFGs of both the application's functions and the code being searched for,
    which can significantly change with software version or obfuscation.
    \framework is resilient to such differences as long as the function's semantics
    remain the same. Unfortunately, we were also unable to obtain the source code
    or detailed results for comparison.

    DyCLINK~\cite{code_relatives} use dynamic analysis to compute a dependency graph
    between instructions executed during developer supplied unit tests. Code
    similarity is determined by computing an isomorphism between sub graphs, using
    edit distance between PageRank~\cite{pagerank} vectors. DyCLINK targets Java
    applications, and thus, we cannot compare \sysname against it.
    DyCLINK considers methods as similar if they share
    \emph{any} sufficiently similar behavior for a given input, an event much more
    prevalent in C binaries than Java binaries. Many dissimilar C functions behave
    similarly when handling errors (i.e., returning $-1$ on invalid input), while
    Java often favors raising different exceptions based on the specific error
    condition. We, therefore, believe that the common error handling technique in C
    would significantly affect DyCLINK's precision.  \sysname is able to distinguish
    between functions with similar functionality, because the decision tree, which
    encodes semantic similarity, is generated using \emph{differences} in behavior.


    Recently, neural networks have been used in binary analysis.
    Zuo et al.~\cite{zuo2019neural}, trained a neural network to determine
    cross-architecture semantics of basic blocks.
    Liu et al.~\cite{alphadiff}, employ a deep neural network to extract features
    from functions and the binary call graph.
    These features are then used to create a distance metric for determining
    binary similarity.
    Xu et al.~\cite{xu2017neural} use a neural network to compute the embedding of
    a function's CFG to accelerate similarity computation.
    These approaches show promise in improving computer security by utilizing
    research from other research areas.


    \section{Conclusion} \label{sec:concl}

    We introduce \framework, a binary analysis framework that is architecture
    and compilation environment agnostic.
    Instead of measuring code properties, \framework abstracts functions into
    sets of input and output program states,
    information guaranteed to be stable across compilation environments.
    Our proof-of-concept implementation, \sysname, has a high \accuracy accuracy
    when identifying functions in binaries generated from various
    configurations, remains highly accurate even against purposefully
    obfuscated code, and can generalize to other architectures with minimal
    effort.
    We also show that the \sysname can readily scale to large binaries.
    We will release \sysname as open source upon acceptance.

    \bibliographystyle{IEEEtran}
    \bibliography{bib}

\end{document}